\documentclass{ws-rv9x6}
\usepackage[square]{ws-rv-van}
\usepackage{bm}

\renewcommand*\thesection{\arabic{section}}

\renewcommand{\thefigure}{\arabic{figure}}

\newcommand{\kf}{k_{\rm F}}
\newcommand{\spin}{{\bm \sigma}_1 \cdot {\bm \sigma}_2}
\newcommand{\ef}{\varepsilon_{{\rm F}\uparrow}}
\newcommand{\mev}{\, \text{MeV}}

\begin{document}

\chapter*{Three-body interactions in Fermi systems}

\author[B. Friman and A. Schwenk]{B. Friman$^{a,}$\footnote{b.friman@gsi.de}
and A. Schwenk$^{b,c,}$\footnote{schwenk@physik.tu-darmstadt.de}}

\address{$^a$GSI Helmholtzzentrum f\"ur Schwerionenforschung GmbH, \\
64291 Darmstadt, Germany \\
$^b$ExtreMe Matter Institute EMMI, \\
GSI Helmholtzzentrum f\"ur Schwerionenforschung GmbH, \\
64291 Darmstadt, Germany \\
$^c$Institut f\"ur Kernphysik, Technische Universit\"at Darmstadt, \\
64289 Darmstadt, Germany}

\begin{abstract}
We show that the contributions of three-quasiparticle interactions
to normal Fermi systems at low energies and temperatures are
suppressed by $n_q/n$ compared to two-body interactions, where $n_q$
is the density of excited or added quasiparticles and $n$ is the
ground-state density. For finite Fermi systems, three-quasiparticle
contributions are suppressed by the corresponding ratio of particle
numbers $N_q/N$. This is illustrated for polarons in strongly
interacting spin-polarized Fermi gases and for valence neutrons in
neutron-rich calcium isotopes.
\end{abstract}

\body

\section{Introduction}

Three-nucleon (3N) forces and advancing microscopic many-body methods
are a frontier in the physics of nuclei and nucleonic matter in stars.
New facets of 3N forces are revealed in neutron-rich nuclei, such as
their role in determining the location of the neutron
dripline~\cite{oxygen,CCoxygen} and in elucidating the doubly-magic
nature of $^{48}$Ca~\cite{calcium}. Three- and higher-body forces are
also the dominant uncertainty in constraining the properties of
neutron-rich matter at nuclear densities and thus the structure of
neutron stars~\cite{Kai,Stefano}. At the same time, 3N forces are at
the center of developing systematic interactions based on effective
field theory (EFT)~\cite{RMP}, thus linking the nuclear forces
frontier with the experimental exploration of neutron-rich nuclei.

For light nuclei, ab-initio methods have unambiguously established the
quantitative role of 3N forces for ground-state properties,
excitations and reactions~\cite{GFMC,NCSM}. These results are
consistent with estimates of 3N force contributions $\langle V_{\rm
3N} \rangle \sim 0.1 - 1 \mev$ per triplet in few-nucleon
systems~\cite{Friar}. However, this scaling with the number of
triplets, which would suggest that 3N forces become more important
than two-nucleon (NN) interactions with increasing particle number
(independent of the density), has to break down for larger
nuclei. This is because nuclear forces have a finite range~$R$
(pion-exchange or shorter), so that in larger systems the interactions
of three particles are restricted to volumes $R^3 < L^3$, where $L$ is
the size of the nucleus. As a result, in nuclear matter 3N forces
scale with the density $n$ (and not the number of triplets to pairs)
compared to two-nucleon interactions $\langle V_{\rm 3N} \rangle \sim
n \, R^3 \, \langle V_{\rm NN} \rangle$ (compare, e.g., the results in
Ref.~\cite{nucmatt}).

For larger nuclei, three-body interactions can be classified according
to a finite-density reference state. In Fermi liquid theory, this
corresponds to the interacting ground state, which is often taken to
be a core nucleus. While 3N force contributions to the energy of the
core nucleus are important, at the level of accuracy of present
calculations for medium-mass nuclei, there is no evidence for residual
three-body interactions between {\it valence} nucleons. For example, a
recent analysis concludes~\cite{Talmi}:~``So far, no evidence was
found for the effects of three-body interactions on states of valence
nucleons. In case where rather pure shell-model configurations were
observed, states and energies were well determined by effective
two-body interactions''. Clearly this must depend on the number of
particles in the core, because in light nuclei with only valence
particles, three-body interactions are significant. In the framework
of Fermi liquid theory, this suggests that three-quasiparticle
interactions are small at low energies, both for excitations and when
valence nucleons are added.

In this paper, we discuss the impact of three-quasiparticle
interactions in normal Fermi systems. These questions have been
touched on briefly in the literature~\cite{Brandow,Chris}. For
example, Brandow writes:~``The weakness of the apparent three-body
effects is the essential content of the statement that a Fermi liquid
may be viewed as a low-density gas of weakly interacting
quasiparticles''. Considering the developments for 3N forces, it is
however important to revisit these issues. After an introduction to
Fermi liquid theory, we show in Section~\ref{3qp} that the
contributions of three-quasiparticle interactions to normal Fermi
systems at low energies and temperatures are suppressed by $n_q/n$
compared to two-body interactions, where $n_q$ is the density of
excited or added quasiparticles and $n$ is the ground-state density,
or by the corresponding ratio of particle numbers $N_q/N$. This
explains why the shell model with effective two-body interactions
works so well. It also enables us to estimate at which level residual
three-body interactions are expected to contribute. These results
demonstrate that there is a change from few-body systems and light
nuclei to normal Fermi systems. For larger nuclei and nucleonic
matter, the contributions from residual three-body interactions are
small when the system is weakly excited (including excitations where
valence nucleons are added), even if 3N forces are significant for the
interacting ground state (the core nucleus).

\vspace*{4mm}

\noindent
We dedicate this paper to Gerry Brown on the occasion of his 85th
birthday. Gerry is one of the pioneers in the theory and microscopic
understanding of Fermi systems. We were extremely fortunate to start
on many-body problems and Fermi liquid theory with him as a teacher
and mentor when we were students in Stony Brook. Using Gerry's words
in ``Fly with eagles''~\cite{eagles}, he has been an eagle for
generations of nuclear theorists and certainly for both of us.

\section{Fermi liquid theory as an effective theory}
\label{FLT}

Much of our understanding of strongly interacting Fermi systems at low
energies and temperatures goes back to the seminal work of Landau in
the late fifties~\cite{Landau1,Landau2,Landau3}. Landau was able to
express macroscopic observables in terms of microscopic properties of
the elementary excitations, the so-called quasiparticles, and their
residual interactions. In order to illustrate Landau's arguments here,
we consider a uniform system of non-relativistic spin-$1/2$ fermions
at zero temperature.

Landau assumed that the low-energy, elementary excitations of the
interacting system can be described by effective degrees of freedom,
the quasiparticles. Due to translational invariance, the states of the
uniform system are eigenstates of the momentum operator. The quasiparticles
are much like single-particle states in the sense that for each
momentum there is a well-defined quasiparticle energy.\footnote{In
general, a quasiparticle state is not an energy eigenstate, but rather
a resonance with a non-zero width. For quasiparticles close to the
Fermi surface, the width is small and the corresponding life-time is
large; hence the quasiparticle concept is useful for time scales short
compared to the quasiparticle life-time.} Landau assumed that there
is a one-to-one correspondence between the quasiparticles and the
single-particle states of a free Fermi gas. For a superfluid system,
this one-to-one correspondence does not exist, and Landau's theory
must be suitably modified, as discussed by Larkin and
Migdal~\cite{LarkinMigdal} and Leggett~\cite{Leggett}.

The one-to-one correspondence starts from a free Fermi gas consisting
of $N$ particles, where the ground state is given by a filled Fermi
sphere in momentum space. The particle number density $n$ and the
ground-state energy $E_0$ are given by (with $\hbar=c=1$)
\begin{equation}
n = \frac{1}{V} \sum_{\mathbf{p} \sigma} n_{\mathbf{p} \sigma}^0 
= \frac{\kf^3}{3 \pi^2} \quad \text{and} \quad
E_0 = \sum_{\mathbf{p} \sigma} 
\frac{\mathbf{p}^2}{2 m} \, n_{\mathbf{p} \sigma}^0
= \frac{3}{5} \frac{\kf^2}{2m} \, N \,,
\end{equation}
where $\kf$ denotes the Fermi momentum, $V$ the volume, and
$n_{\mathbf{p} \sigma}^0 = \theta(\kf - |\mathbf{p}|)$ is the Fermi-Dirac
distribution function at zero temperature for particles with momentum
$\mathbf{p}$, spin projection $\sigma$, and mass $m$.
By adding particles or holes, the distribution function is changed by 
$\delta n_{\mathbf{p}\sigma} = n_{\mathbf{p}\sigma}-n_{\mathbf{p}\sigma}^0$, 
and the total energy of the system by
\begin{equation}
\delta E = E - E_0 = \sum_{\mathbf{p} \sigma} \frac{\mathbf{p}^2}{2m} \, 
\delta n_{\mathbf{p} \sigma} \,.
\end{equation}
When a particle is added in the state $\mathbf{p}\sigma$, one has
$\delta n_{\mathbf{p}\sigma} =1$ and when a particle is removed 
(a hole is added) $\delta n_{\mathbf{p}\sigma} =-1$.

In the interacting system the corresponding state is one with a
quasiparticle added or removed, and the change in energy is given by
\begin{equation}
\delta E = \sum_{\mathbf{p} \sigma} \varepsilon_{\mathbf{p} \sigma} \,
\delta n_{\mathbf{p} \sigma} \,,
\end{equation}
where $\varepsilon_{\mathbf{p} \sigma}=\delta E/\delta
n_{\mathbf{p}\sigma}$ denotes the quasiparticle energy. When two or
more quasiparticles are added to the system, an additional term takes
into account the interaction between the quasiparticles:
\begin{equation}
\delta E = \sum_{\mathbf{p} \sigma} \varepsilon^0_{\mathbf{p} \sigma} \, 
\delta n_{\mathbf{p} \sigma} + \frac{1}{2V} 
\sum_{\mathbf{p}_1 \sigma_1, \mathbf{p}_2 \sigma_2} 
f_{\mathbf{p}_1 \sigma_1 \mathbf{p}_2 \sigma_2} \, \delta n_{\mathbf{p}_1 \sigma_1} \, 
\delta n_{\mathbf{p}_2 \sigma_2} \,.
\label{eq:delta_E}
\end{equation}
Here $\varepsilon^0_{\mathbf{p} \sigma}$ is the quasiparticle energy
in the ground state. In Section~\ref{3qp}, we will show that the
expansion in $\delta n$ is general and does not require weak
interactions. The small expansion parameter in Fermi liquid theory is
the density of quasiparticles, or equivalently the excitation energy,
and not the strength of the interaction. This allows a systematic
treatment of strongly interacting systems at low temperatures.

In normal Fermi systems, the quasiparticle concept makes sense only
for states close to the Fermi surface, where the quasiparticle
life-time $\tau_{\mathbf{p}} \sim (p - \kf)^2$ is
long~\cite{BaymPethick}.  In particular, states deep in the Fermi sea,
which are occupied in the ground-state distribution, do not correspond
to well-defined quasiparticles. Accordingly, we refer to the
interacting ground state that corresponds to a filled Fermi sea in the
non-interacting system as a state with no quasiparticles. In a weakly
excited state the quasiparticle distribution $\delta n_{\mathbf{p}
\sigma}$ is generally non-zero only for states close to the Fermi
surface.

The second term in Eq.~(\ref{eq:delta_E}), the quasiparticle
interaction $f_{\mathbf{p}_1 \sigma_1 \mathbf{p}_2 \sigma_2}$, has no
correspondence in the non-interacting Fermi gas. In an
excited state with more than one quasiparticle, the quasiparticle
energy is modified according to
\begin{equation}
\varepsilon_{\mathbf{p} \sigma}
= \frac{\delta E}{\delta n_{\mathbf{p}\sigma}}
= \varepsilon^0_{\mathbf{p} \sigma}+\frac{1}{V}
\sum_{\mathbf{p}_2 \sigma_2} f_{\mathbf{p} \sigma \mathbf{p}_2 \sigma_2} \, \delta 
n_{\mathbf{p}_2 \sigma_2} \,,
\end{equation}
where the changes are effectively proportional to the quasiparticle density.

The quasiparticle interaction can be understood microscopically from
the second variation of the energy with respect to the quasiparticle
distribution,
\begin{equation}
f_{\mathbf{p}_1 \sigma_1 \mathbf{p}_2 \sigma_2} = V \, 
\frac{\delta^2 E}{\delta n_{\mathbf{p}_1 \sigma_1} \delta n_{\mathbf{p}_2 \sigma_2}} 
= V \, \frac{\delta \varepsilon_{\mathbf{p}_1 \sigma_1}}{\delta 
n_{\mathbf{p}_2 \sigma_2}} \,.
\label{eq:f}
\end{equation}
Diagrammatically, this variation corresponds to cutting one of the
fermion lines in a given energy diagram and labeling the incoming and
outgoing fermion by $\mathbf{p}_1 \sigma_1$, followed by a second
variation leading to $\mathbf{p}_2 \sigma_2$. For the uniform system,
the resulting contributions to $f_{\mathbf{p}_1 \sigma_1 \mathbf{p}_2
\sigma_2}$ are quasiparticle reducible in the particle-particle and
in the exchange particle-hole (induced interaction) channels, but
irreducible in the direct particle-hole (zero sound)
channel~\cite{AK,GerryRMP,BB}. The zero-sound-channel reducible diagrams
are generated by the particle-hole scattering equation~\cite{Landau3}.
With Babu, one of Gerry's seminal contributions to Fermi liquid theory
was to derive an integral equation that self-consistently takes into
account induced interactions due to the polarization of the
medium~\cite{BB}. The Babu-Brown induced interaction is still one of
the few non-perturbative approaches for calculating Fermi liquid
parameters that have been implemented in practice.

Landau's theory of normal Fermi liquids is an effective low-energy
theory in the modern sense~\cite{Shankar,Polchinski}. The effective
theory incorporates the symmetries of the system and the low-energy
couplings can be fixed by experiment or calculated microscopically
based on the underlying theory. In an isotropic and spin-saturated
system, such as liquid $^{3}$He, the quasiparticle interaction can be
decomposed as
\begin{equation}
f_{\mathbf{p}_1 \sigma_1 \mathbf{p}_2 \sigma_2} = f_{\mathbf{p}_1 \mathbf{p}_2}^s 
+ f^a_{\mathbf{p}_1 \mathbf{p}_2} \: \spin \,,
\end{equation}
which is the most general form consistent with $SU(2)$ spin
symmetry.\footnote{In nuclear physics the notation $f_{\mathbf{p}_1
\mathbf{p}_2} = f^s_{\mathbf{p}_1 \mathbf{p}_2}$ and
$g_{\mathbf{p}_1 \mathbf{p}_2} = f^a_{\mathbf{p}_1 \mathbf{p}_2}$ is
generally used.}  For nuclear systems, the quasiparticle interaction
includes additional terms that take into account the isospin
dependence and non-central tensor
contributions~\cite{MigdalBook,GerryPR,tensor}. However, for
our discussion here, the spin and isospin dependence is not important.

For the uniform system, Eq.~(\ref{eq:f}) yields the quasiparticle
interaction only for forward scattering (low momentum transfers). In
the particle-hole channel, this corresponds to the long-wavelength
limit. This restriction, which is consistent with considering low
excitation energies, constrains the momenta $\mathbf{p}_1$ and
$\mathbf{p}_2$ to be close to the Fermi surface, $|\mathbf{p}_1| =
|\mathbf{p}_2| = \kf$. The quasiparticle interaction then depends only
on the angle between $\mathbf{p}_1$ and $\mathbf{p}_2$. It is
convenient to expand this dependence on Legendre polynomials
\begin{equation}
f_{\mathbf{p}_1 \mathbf{p}_2}^{s/a} = f^{s/a}(\cos \theta_{\mathbf{p}_1 \mathbf{p}_2}) 
= \sum_l \, f_l^{s/a} \, P_l(\cos \theta_{\mathbf{p}_1 \mathbf{p}_2}) \,,
\label{eq:f_expansion}
\end{equation}
and to define the dimensionless Landau Parameters $F_l^{s/a}$ by
\begin{equation}
F_l^{s/a} = N(0) \, f_l^{s/a} \,,
\end{equation}
where $N(0) = \frac{1}{V} \sum_{\mathbf{p} \sigma} 
\delta(\varepsilon_{\mathbf{p} \sigma} - \mu) = m^* \kf/\pi^2$
denotes the quasiparticle density of states at the Fermi surface.

The Landau parameters can be directly related to macroscopic
properties of the system. Here we mention only the specific heat
$c_V$, which at low temperature is determined by the effective mass
given by $F_1^s$,
\begin{align}
\frac{m^*}{m} &= 1 + \frac{F^s_1}{3} \,, \\
c_V &= \frac{m^* \kf}{3} \, k_B^2 T \,,
\end{align}
and the incompressibility $\kappa$, which is related to $F_0^s$,
\begin{equation}
\kappa = - \frac{9 V}{n} \frac{\partial P}{\partial V} 
= \frac{3 \kf^2}{m^*} \, (1 + F_0^s) \,.
\end{equation}
Fermi liquid theory has been very successful in describing
low-temperature Fermi liquids, in particular liquid
$^3$He~\cite{BaymPethick}. The first applications to nuclear systems
were pioneered by Migdal~\cite{MigdalBook} and first microscopic
calculations for nuclei and nuclear matter by Gerry and collaborators
(for a review, see Ref.~\cite{GerryPR}). Recently, advances using
renormalization group (RG) methods for nuclear forces~\cite{Vlowk} and
Fermi systems~\cite{Shankar} have lead to the development of a
non-perturbative RG approach for nucleonic matter~\cite{RGnm}, to a
first complete study of the spin structure of induced
interactions~\cite{tensor}, and to new calculations of Fermi liquid
parameters~\cite{indint,Kaiser}.

\section{Three-quasiparticle interactions}
\label{3qp}

In Section~\ref{FLT}, we introduced Fermi liquid theory as an
expansion in the density of quasiparticles $\delta n/V$. In
applications of Fermi liquid theory to date, even for liquid $^3$He,
which is a very dense and strongly interacting system, this expansion
is truncated after the second-order $(\delta n)^2$ term, including
only pairwise interactions of quasiparticles. However, for a strongly
interacting system, there is a priori no reason that three-body (or
higher-body) interactions between quasiparticles are small. In this
section, we discuss the convergence of this expansion.
Three-quasiparticle interactions arise from iterated two-body forces,
leading to three- and higher-body clusters in the linked-cluster
expansion, or through many-body forces. While three-body forces play
an important role in nuclear physics~\cite{RMP}, little is known about
them in other Fermi liquids. Nevertheless, in strongly interacting
systems, the contributions of many-body clusters can in general be
significant, leading to potentially important $(\delta n)^3$ terms in
the Fermi liquid expansion, also in the absence of three-body forces:
\begin{equation}
\delta E = \sum_1 \varepsilon^0_1 \: \delta n_1
+ \frac{1}{2V} \sum_{1,2} f^{(2)}_{1,2} \: \delta n_1 \, \delta n_2
+ \frac{1}{6V^2} \sum_{1,2,3} f^{(3)}_{1,2,3} \: \delta n_1 \, \delta n_2
\, \delta n_3 \,.
\label{eq:delta_E3}
\end{equation}
Here $f^{(n)}_{1,\ldots,n}$ denotes the $n$-quasiparticle interaction
(the Landau interaction is $f \equiv f^{(2)}$) and we have introduced
the short-hand notation $n \equiv {\bf p}_n, \sigma_n$.

In order to better understand the expansion, Eq.~(\ref{eq:delta_E3}),
around the interacting ground state with $N$ fermions, consider
exciting or adding $N_q$ quasiparticles with $N_q \ll N$. The
microscopic contributions from many-body clusters or from many-body
forces can be grouped into diagrams containing zero, one, two, three,
or more quasiparticle lines.  The terms with zero quasiparticle lines
contribute to the interacting ground state for $\delta n = 0$, whereas
the terms with one, two, and three quasiparticle lines contribute to
$\varepsilon^0_1$, $f^{(2)}_{1,2}$, and $f^{(3)}_{1,2,3}$,
respectively (these also depend on the ground-state density due to the
$N$ fermion lines). The terms with more than three quasiparticle lines
would contribute to higher-quasiparticle interactions. Because a
quasiparticle line replaces a line with $N$ fermions when going from
$\varepsilon^0_1$ to $f^{(2)}_{1,2}$, and from $f^{(2)}_{1,2}$ to
$f^{(3)}_{1,2,3}$, it is intuitively clear that the contributions due
to three-quasiparticle interactions are suppressed by $N_q/N$ compared
to two-quasiparticle interactions, and that the Fermi liquid expansion
is effectively an expansion in $N_q/N$ or
$n_q/n$~\cite{PinesNozieres}. This will be discussed in detail and
illustrated with examples in the following sections.

\subsection{General considerations}

Fermi liquid theory applies to normal Fermi systems at low energies
and temperatures, or equivalently at low quasiparticle densities. We
first consider excitations that conserve the net number of quasiparticles,
$\delta N = \sum_{\mathbf{p} \sigma} \delta n_{\mathbf{p} \sigma}=0$,
so that the number of quasiparticles equals the number of
quasiholes. This corresponds to the lowest energy excitations of
normal Fermi liquids. We denote their energy scale by
$\Delta$. Excitations with one valence particle or quasiparticle added
start from energies of order the chemical potential $\mu$. In the case
of $\delta N=0$, the contributions of two-quasiparticle interactions
are of the same order as the first-order $\delta n$ term, but
three-quasiparticle interactions are suppressed by
$\Delta/\mu$~\cite{Chris}. This is the reason that Fermi liquid theory
with only two-body Landau parameters is so successful in describing
even strongly interacting and dense Fermi liquids. This counting is
best seen from the variation of the free energy $F=E-\mu N$,
\begin{align}
\delta F &= \delta(E-\mu N) \nonumber \\
&= \sum_1 (\varepsilon^0_1 - \mu) \, \delta n_1
+ \frac{1}{2V} \sum_{1,2} f^{(2)}_{1,2} \: \delta n_1 \, \delta n_2
+ \frac{1}{6V^2} \sum_{1,2,3} f^{(3)}_{1,2,3} \: \delta n_1 \, \delta n_2
\, \delta n_3 \,,
\label{eq:delta_F3}
\end{align}
which for $\delta N=0$ is equivalent to $\delta E$ of
Eq.~(\ref{eq:delta_E3}). The quasiparticle distribution
is $|\delta n_{\mathbf{p} \sigma}| \sim 1$ within a shell around the
Fermi surface $|\varepsilon^0_{\mathbf{p} \sigma} - \mu| \sim
\Delta$. The first-order $\delta n$ term is therefore proportional to
$\Delta$ times the number of quasiparticles
$\sum_{\mathbf{p} \sigma} |\delta n_{\mathbf{p} \sigma}| = N_q
\sim N (\Delta/\mu)$,
\begin{equation}
\sum_1 (\varepsilon^0_1 - \mu) \, \delta n_1 \sim \frac{N
\Delta^2}{\mu} \,.
\end{equation}
Correspondingly, the contribution of two-quasiparticle interactions
yields
\begin{equation}
\frac{1}{2V} \sum_{1,2} f^{(2)}_{1,2} \: \delta n_1 \, \delta n_2
\sim \frac{1}{V} \, \langle f^{(2)} \rangle \, \biggl(\frac{N 
\Delta}{\mu}\biggr)^2 
\sim \langle F^{(2)} \rangle \, \frac{N \Delta^2}{\mu} \,,
\label{eq:f2scaling}
\end{equation}
where $\langle F^{(2)} \rangle = n \, \langle f^{(2)} \rangle /\mu$ is
an average dimensionless coupling on the order of the Landau
parameters. Even in the strongly interacting, scale-invariant case
(see Section~\ref{polaron}) $\langle f^{(2)} \rangle \sim 1/\kf$;
hence $\langle F^{(2)} \rangle \sim 1$ and the contribution of
two-quasiparticle interactions is of the same order as the first-order
term. However, the three-quasiparticle contribution is of order
\begin{equation}
\frac{1}{6V^2} \sum_{1,2,3} f^{(3)}_{1,2,3} \: \delta n_1 \, \delta n_2
\, \delta n_3 
\sim \frac{n^2}{\mu} \, \langle f^{(3)} \rangle \, \frac{N \Delta^3}{\mu^2}
\sim \langle F^{(3)} \rangle \,\frac{N \Delta^3}{\mu^2} \,.
\label{eq:f3scaling}
\end{equation}
Therefore at low excitation energies this is suppressed by
$\Delta/\mu$, compared to two-quasiparticle interactions, even if the
dimensionless three-quasiparticle interaction $\langle F^{(3)} \rangle
= n^2 \langle f^{(3)} \rangle /\mu$ is strong (of order 1). Similarly,
higher $n$-body interactions are suppressed by
$(\Delta/\mu)^{n-2}$. Normal Fermi systems at low energies are weakly
coupled in this sense.  The small parameter is the ratio of the
excitation energy per particle to the chemical potential. These
considerations hold for {\it all} normal Fermi systems where the
underlying interparticle interactions are finite range.

The Fermi liquid expansion in $\Delta/\mu$ is equivalent to an
expansion in $N_q/N \sim \Delta/\mu$, the ratio of the number of
quasiparticles and quasiholes $N_q$ to the number of particles $N$ in
the interacting ground state, or an expansion in the density of
excited quasiparticles over the ground-state density, $n_q/n$.

We conclude this section with a discussion of the expansion for the
energy $\delta E$ given by Eq.~(\ref{eq:delta_E3}), for the case where
$N_q$ quasiparticles or valence particles are added to a Fermi-liquid
ground state. In this case, $\delta N \neq 0$ and the first-order
term is
\begin{equation}
\sum_1 \varepsilon^0_1 \: \delta n_1 \sim \mu N_q 
\sim \mu \, \frac{N \Delta}{\mu}
\sim N \Delta \,,
\end{equation}
while the contribution of two-quasiparticle interactions is suppressed
by $N_q/N \sim \Delta/\mu$ and that of three-quasiparticle
interactions by $(N_q/N)^2$. Therefore, either for $\delta N=0$ or
$\delta N \neq 0$, the contributions of three-quasiparticle
interactions are suppressed for normal Fermi systems at low excitation
energies. In the following sections, we will illustrate this for
polarons in strongly interacting spin-polarized Fermi gases and for
valence neutrons in neutron-rich calcium isotopes.

\subsection{Strongly interacting spin-polarized Fermi gases}
\label{polaron}

Experiments with spin-polarized Fermi
gases~\cite{Zwierlein,Hulet,Shin,MITpolaron,Paris} enable a unique
exploration of strongly interacting Fermi systems and universal
properties. We consider a system with two spin states and large S-wave
scattering length interactions. In the limit of extreme population
imbalance, the physics is governed by a single spin-down fermion
interacting strongly with the spin-up Fermi sea. This spin-down
fermion forms a quasiparticle, the so-called Fermi
polaron~\cite{Chevyreview}, with energy $E_p$ and effective mass
$m_p^*$. Polarons have been directly observed in cold atomic
gases using rf spectroscopy~\cite{MITpolaron}.

For large scattering lengths, $1/a_{\rm s}=0$, the polaron energy is
universal (it depends only on the density of spin-up fermions): $E_p =
\eta \, \ef$, where $\ef = (6 \pi^2 n_\uparrow)^{2/3}/(2m)$ is the
spin-up Fermi energy~\cite{Chevy}. The polaron binding $\eta = -0.615$
and effective mass $m_p^*/m = 1.20(2)$ have been determined using
Monte-Carlo methods~\cite{PS} and are in excellent agreement with
experiment~\cite{Paris}. The polaron energy constrains the energy gain
for large asymmetries in the Fermi liquid phase of spin-polarized
Fermi gases. This determines the critical polarization for
superfluidity and sets limits on the phase diagram and the existence
of partially-polarized phases~\cite{Chevy,BF,Shin}.

We can expand the strongly interacting spin-polarized Fermi gas around
the fully polarized system with $N_\uparrow$ particles by adding $N_q
= N_\downarrow \ll N_\uparrow$ polarons. Following
Eq.~(\ref{eq:delta_E3}), the change in the energy density is given
by~\cite{Lobo}
\begin{equation}
\frac{\delta E_\downarrow}{V} =  \varepsilon^0_1 \, n_\downarrow
+ \frac{1}{2} \, f^{(2)} \, n_\downarrow^2
+ \frac{1}{6} \, f^{(3)} \, n_\downarrow^3 \,,
\end{equation}
where $\varepsilon^0_1$ is the average quasiparticle energy with
contributions from both the polaron binding and the kinetic energy
(with the single-polaron effective mass),
\begin{equation}
\varepsilon^0_1 = \eta \, \ef + \frac{3}{5} \, 
\frac{(6 \pi^2 n_\downarrow)^{2/3}}{2 m_p^*} \,.
\end{equation}
The average two-quasiparticle interaction $f^{(2)}$ is due to induced
interactions mediated by the spin-up Fermi sea~\cite{polaronind} and
scales with the Fermi energy and the density of spin-up fermions,
\begin{equation}
f^{(2)} = \frac{\ef}{n_\uparrow} \, F^{(2)} \,,
\end{equation}
and correspondingly for the three-quasiparticle interaction,
\begin{equation}
f^{(3)} = \frac{\ef}{n_\uparrow^2} \, F^{(3)} \,.
\end{equation}
For large scattering lengths, $1/a_{\rm s}=0$, the only scale is set
by the spin-up density, and therefore the average $F^{(2)}$ and
$F^{(3)}$ are dimensionless constants. In general, the two- and
three-quasiparticle interactions also depend on the angles between the
quasiparticles close to the Fermi surface (in particular, the
effective mass at finite polaron density is given by an appropriately
defined $l=1$ Landau parameter~\cite{Sjoberg}), but for the general
estimates here, we can consider an average interaction relevant for
the energy contribution. If additional scales are significant, such as
the effective range or other ranges~$R$, $F^{(2)}$ and $F^{(3)}$ will
depend on $R^3 n_\uparrow$. Monte-Carlo calculations of $F^{(2)}$
give $F^{(2)} = 6 B/5 \approx 0.17$ ($B \approx 0.14$ in
Ref.~\cite{Pilati}), which is small (compared to the normal symmetric
phase) due to the Pauli principle for spin-down fermions. These scaling
results for large scattering lengths demonstrate nicely the
suppression of three-quasiparticle contributions by $(F^{(3)}/F^{(2)})
(n_\downarrow/n_\uparrow)$, in line with the results of the previous
section.

\subsection{Neutron-rich nuclei}

\begin{figure}[t]
\begin{center}
\includegraphics[scale=0.65,clip=]{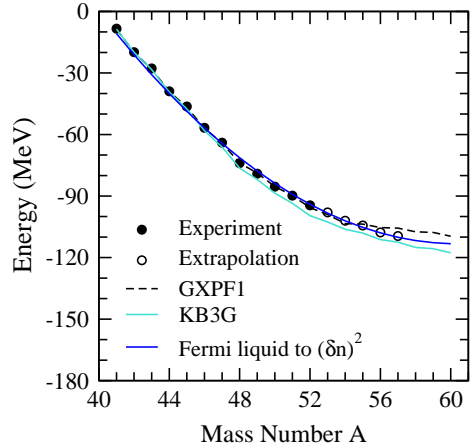}
\end{center}
\caption{Ground-state energies of calcium isotopes relative to
$^{40}$Ca as a function of mass number $A$, taken from the AME2003
atomic mass evaluation~\cite{AME2003} based on
experimental energies to $^{52}$Ca and an
extrapolation from $^{53}$Ca to $^{58}$Ca (the heaviest neutron-rich
calcium isotopes known to exist~\cite{53-58Ca}). In addition,
we show the energies obtained from shell-model calculations with
phenomenological two-body interactions KB3G~\cite{KB3G} and
GXPF1~\cite{GXPF1}, and the fit using the Fermi liquid
expansion, Eq.~(\ref{FLfinite}), to second order.\label{ca}}
\end{figure}

Next we illustrate the suppression of three-quasiparticle terms for
finite Fermi systems. We consider valence neutrons in neutron-rich
calcium isotopes, where the interacting ground state is taken to be
the $^{40}$Ca core.\footnote{The properties of medium-mass (and
heavier) nuclei are often also calculated in energy-density
functional approaches, where particle-particle (pair) correlations
are included by generalizing the ground state to a
(particle-number-projected) superfluid ground state. Figure~\ref{ca}
shows that pairing effects responsible for the
odd-even-mass-staggering are relatively weak in nuclei.} This is
also an interesting system, because recent calculations (with
empirical single-particle or quasiparticle energies) have shown that
the dominant contributions from chiral 3N forces are due to
interactions between two valence neutrons and one core
nucleon~\cite{calcium}. This corresponds to the normal-ordered
two-body part of 3N forces, which is enhanced by the number of core
nucleons. In the language of Fermi liquid theory, these are 3N force
contributions to the two-quasiparticle interaction.

As shown in Fig.~\ref{ca}, the ground-state energies of neutron-rich
calcium isotopes are well reproduced with effective two-body
interactions in the shell model~\cite{KB3G,GXPF1}. The differences
between the phenomenological interactions are however amplified with
increasing neutron number.

For a finite system, the Fermi liquid expansion,
Eq.~(\ref{eq:delta_E3}), is given by:
\begin{equation}
\delta E \sim \mu \, \biggl[ N_q + \langle F^{(2)} \rangle \, 
\frac{N_q(N_q-1)}{2 A} + \langle F^{(3)} \rangle \, 
\frac{N_q(N_q-1)(N_q-2)}{6 A^2} \biggr] \,,
\label{FLfinite}
\end{equation}
where we have used $\varepsilon^0_1 \sim \mu$ and the dimensional
scaling of $f^{(2)}_{12} \sim \mu/n$ and $f^{(3)}_{123} \sim \mu/n^2$
[see Eqs.~(\ref{eq:f2scaling}) and~(\ref{eq:f3scaling})]. In this
example, $N_q$ is the number of valence neutrons and $A=40$ the number
of core nucleons. A fit to the ground-state energies of the AME2003
atomic mass evaluation~\cite{AME2003} including only two-quasiparticle
interactions yields $\mu = -10.8 \mev$ and $\langle F^{(2)} \rangle =
- 2.0$. Overall a very good description of the energies is obtained,
and the fit from $^{41}$Ca to $^{58}$Ca is not sensitive to
three-quasiparticle interactions. The value for $\mu$ is consistent
with typical single-particle energies in $^{41}$Ca and $\langle
F^{(2)} \rangle \sim 1$ is expected for nuclear interactions. We could
improve the description further by accounting for the dependence of
the two-body interaction on the single-particle orbitals (as in the
shell model), instead of using an average $\langle F^{(2)} \rangle$.

Moreover, the Fermi liquid expansion provides an estimate of the
contribution from three-quasiparticle interactions. With $\langle
F^{(3)} \rangle \sim 1$, which is likely to be an overestimate,
because three-neutron interactions are suppressed by the Pauli
principle, the corresponding energy contributions to $^{52}$Ca and
$^{58}$Ca ($N_q=12$ and $18$) are $\delta E \sim 1.5$ and $5.5
\mev$. This is a factor 2 smaller than the spread of the shell-model
results in Fig.~\ref{ca}. However, due to the uncertainty in the
strength of $\langle F^{(3)} \rangle$, microscopic calculations or
more global shell-model analyses of $\langle F^{(3)} \rangle$ are
important to improve this estimate. Finally, the convergence of the
Fermi liquid expansion is improved and the suppression of
three-quasiparticle contributions is even more effective in heavier
nuclei.

\section{Concluding remarks}

We have shown that for normal Fermi systems at low excitation energies
the contributions of three-quasiparticle interactions are suppressed
by the ratio of the quasiparticle density to the ground-state density,
or equivalently by the ratio of the excitation energy over the
chemical potential. This holds for excitations that conserve the
number of particles (excited states of the interacting ground state)
as well as for excitations that add or remove valence particles. This
suppression is general and applies to strongly interacting systems
even with strong, but finite-range three-body forces. However, this
does not imply that the contributions from 3N forces to the
interacting ground-state energy (the energy of the core nucleus in the
context of shell-model calculations), to quasiparticle energies
(single-particle energies), or to two-quasiparticle interactions
(effective two-body interactions) are small. The argument only applies
to the effects of residual three-body interactions at low energies.

The Fermi liquid suppression of three-quasiparticle interactions can
be tested in large-scale shell model calculations, and with advances
in ab-initio methods for larger nuclei, in no-core shell model
calculations with a core~\cite{coreNCSM}, in coupled-cluster
theory~\cite{CC} and with nuclear lattice simulations~\cite{lattice}.
For interparticle interactions where a finite-density reference state
is close to the interacting ground state, the Fermi liquid expansion
also implies that normal-ordered three-body interactions are small.
This can explain why, for low-momentum interactions~\cite{Vlowk},
calculations of nuclei~\cite{CC3N,IMSRG} and nucleonic
matter~\cite{nucmatt,chiralnm} at the normal-ordered two-body level
are so successful.

\section*{Acknowledgments}

We are deeply grateful to Gerry for his very personal and continuous
support and for the scientific inspiration in physics, through
numerous discussions, and in the unique atmosphere he created in Stony
Brook. Without his tremendous investment in people, we would not be
where we are today.

\vspace*{4mm}

\noindent
We also thank R.\ J.\ Furnstahl and C.\ J.\ Pethick for useful
discussions on these topics. This work was supported in part by the
Helmholtz Alliance Program of the Helmholtz Association, contract
HA216/EMMI ``Extremes of Density and Temperature: Cosmic Matter in the
Laboratory'' and the DFG through grant SFB 634.


\begin{thebibliography}{99}
\bibitem{oxygen} T.\ Otsuka, T.\ Suzuki, J.\ D.\ Holt, A.\ Schwenk and 
Y.\ Akaishi, Phys.\ Rev.\ Lett.\ {\bf 105}, 032501 (2010).
\bibitem{CCoxygen} G.\ Hagen, T.\ Papenbrock, D.\ J.\ Dean,
M.\ Hjorth-Jensen and B.\ Velamur Asokan, Phys.\ Rev.\ C {\bf 80},
021306 (2009).
\bibitem{calcium} J.\ D.\ Holt, T.\ Otsuka, A.\ Schwenk and T.\ Suzuki,
arXiv:1009.5984.
\bibitem{Kai} K.\ Hebeler, J.\ M.\ Lattimer, C.\ J.\ Pethick and
A.\ Schwenk, Phys.\ Rev.\ Lett.\ {\bf 105}, 161102 (2010).
\bibitem{Stefano} S.\ Gandolfi, J.\ Carlson and S.\ Reddy, arXiv:1101.1921.
\bibitem{RMP} E.\ Epelbaum, H.-W.\ Hammer and U.-G.\ Mei{\ss}ner,
Rev.\ Mod.\ Phys.\ {\bf 81}, 1773 (2009).
\bibitem{GFMC} S.\ C.\ Pieper, Riv.\ Nuovo Cim.\ {\bf 031}, 709 (2008).
\bibitem{NCSM} P.\ Navr\'{a}til, S.\ Quaglioni, I.\ Stetcu and
B.\ R.\ Barrett, J.\ Phys.\ G {\bf 36}, 083101 (2009).
\bibitem{Friar} J.\ L.\ Friar, ``Nuclear Scales'' in {\it Nuclear Physics
with Effective Field Theory}, Ed.\ R.\ Seki, U.\ van Kolck and M.\
J.\ Savage (World Scientific, Singapore, 1998); nucl-th/9804010.
\bibitem{nucmatt} S.\ K.\ Bogner, A.\ Schwenk, R.\ J.\ Furnstahl and
A.\ Nogga, Nucl.\ Phys.\ A {\bf 763}, 59 (2005); K.\ Hebeler, S.\ K.\ 
Bogner, R.\ J.\ Furnstahl, A.\ Nogga and A.\ Schwenk, arXiv:1012.3381.
\bibitem{Talmi} P.\ van Isacker and I.\ Talmi, Europhys.\ Lett.\ {\bf
90}, 32001 (2010).
\bibitem{Brandow} B.\ H.\ Brandow, Rev.\ Mod.\ Phys.\ {\bf 39}, 771
(1967).
\bibitem{Chris} C.\ J.\ Pethick, ``Selected Topics in the Theory of
Normal Fermi Liquids'', in {\it Lectures in Theoretical Physics},
Ed.\ K.\ T.\ Mahantappa and W.\ E.\ Brittin (Gordon and Breach, New
York, 1969), p. 187.
\bibitem{eagles} G.\ E.\ Brown, Annu.\ Rev.\ Nucl.\ Part.\ Sci.\
{\bf 51}, 1 (2001).
\bibitem{Landau1} L.\ D.\ Landau, Sov.\ Phys.\ JETP {\bf 3}, 920 (1957).
\bibitem{Landau2} L.\ D.\ Landau, Sov.\ Phys.\ JETP {\bf 5}, 101 (1957).
\bibitem{Landau3} L.\ D.\ Landau, Sov.\ Phys.\ JETP {\bf 8}, 70 (1959).
\bibitem{LarkinMigdal} A.\ Larkin and A.\ B.\ Migdal, Sov.\ Phys.\ JETP 
{\bf 17}, 1146 (1963).
\bibitem{Leggett} A.\ J.\ Leggett, Phys.\ Rev.\ A {\bf 140}, 1869 (1965);
{\it ibid.} {\bf 147}, 119 (1966).
\bibitem{BaymPethick} G.\ Baym and C.\ J.\ Pethick, {\it Landau Fermi Liquid
Theory: Concepts and Applications} (Wiley, New York, 1991).
\bibitem{AK} A.\ A.\ Abrikosov and I.\ M.\ Khalatnikov, Rept.\ Prog.\ 
Phys.\ {\bf 22}, 329 (1959).
\bibitem{GerryRMP} G.\ E.\ Brown, Rev.\ Mod.\ Phys.\ {\bf 43}, 1 (1971).
\bibitem{BB} S.\ Babu and G.\ E.\ Brown, Ann.\ Phys.\ {\bf 78} (1973) 1.
\bibitem{Shankar} R.\ Shankar, Rev.\ Mod.\ Phys.\ \textbf{66}, 129 (1994).
\bibitem{Polchinski} J.\ Polchinski, ``Effective Field Theory and the
Fermi Surface'', in {\it Proceedings of the 1992 Theoretical Advanced
Studies Institute in Elementary Particle Physics}, Ed.\ J.\ Harvey and 
J.\ Polchinski (World Scientific, Singapore, 1993), hep-th/9210046.
\bibitem{MigdalBook} A.\ B.\ Migdal, {\it Theory of Finite Fermi Systems
and Applications to Atomic Nuclei} (Interscience, New York, 1967).
\bibitem{GerryPR} S.-O.\ B\"ackman, G.\ E.\ Brown and J.\ Niskanen,
Phys.\ Rept. {\bf 124}, 1 (1985).
\bibitem{tensor} A.\ Schwenk and B.\ Friman, Phys.\ Rev.\ Lett.\ {\bf 92},
082501 (2004).
\bibitem{Vlowk} S.\ K.\ Bogner, R.\ J.\ Furnstahl and
A.\ Schwenk, Prog.\ Part.\ Nucl.\ Phys. {\bf 65}, 94 (2010).
\bibitem{RGnm} A.\ Schwenk, B.\ Friman and G.\ E.\ Brown, Nucl.\ Phys.\ 
A {\bf 713}, 191 (2003).
\bibitem{indint} A.\ Schwenk, G.\ E.\ Brown and B.\ Friman, Nucl.\
Phys.\ A {\bf 703}, 745 (2002).
\bibitem{Kaiser} N.\ Kaiser, Nucl.\ Phys.\ A {\bf 768}, 99 (2006).
\bibitem{PinesNozieres} D.\ Pines and P.\ Nozi\`eres, {\it The Theory of 
Quantum Liquids} (Volume 1, Advanced Book Classics, Westview Press, 1999).
\bibitem{Zwierlein} M.\ W.\ Zwierlein, A.\ Schirotzek, C.\ H.\ Schunck
and W.\ Ketterle, Science {\bf 311}, 492 (2006); Y.\ Shin, M.\ W.\
Zwierlein, C.\ H.\ Schunck, A.\ Schirotzek and W.\ Ketterle, Phys.\
Rev.\ Lett.\ {\bf 97}, 030401 (2006).
\bibitem{Hulet} G.\ B.\ Partridge, W.\ Li, R.\ I.\ Kamar, Y.-a.\ Liao
and R.\ G.\ Hulet, Science {\bf 311}, 503 (2006).
\bibitem{Shin} Y.-i.\ Shin, C.\ H.\ Schunck, A.\ Schirotzek and 
W.\ Ketterle, Nature {\bf 451}, 689 (2008).
\bibitem{MITpolaron} A.\ Schirotzek, C.-H.\ Wu, A.\ Sommer and M.\
Zwierlein, Phys.\ Rev.\ Lett.\ {\bf 102}, 230402 (2009).
\bibitem{Paris} S.\ Nascimb\`ene, N.\ Navon, K.\ J.\ Jiang, F.\ Chevy
and C.\ Salomon, Nature {\bf 463}, 1057 (2010).
\bibitem{Chevyreview} F.\ Chevy and C.\ Mora, Rep.\ Prog.\ Phys.\
{\bf 73}, 112401 (2010).
\bibitem{Chevy} F.\ Chevy, Phys.\ Rev.\ A {\bf 74}, 063628 (2006).
\bibitem{PS} N.\ Prokof'ev and B.\ Svistunov, Phys.\ Rev.\ B {\bf 77},
020408(R) (2008); {\it ibid.} {\bf 77}, 125101 (2008).
\bibitem{BF} A.\ Bulgac and M.\ M.\ Forbes, Phys.\ Rev.\ A {\bf 75},
031605(R) (2007).
\bibitem{Lobo} C.\ Lobo, A.\ Recati, S.\ Giorgini and S.\ Stringari,
Phys.\ Rev.\ Lett.\ {\bf 97}, 200403 (2006).
\bibitem{polaronind} C.\ Mora and F.\ Chevy, Phys.\ Rev.\ Lett.\
{\bf 104}, 230402 (2010); Z.\ Yu, S.\ Z\"ollner and C.\ J.\ Pethick,
Phys.\ Rev.\ Lett.\ {\bf 105}, 188901 (2010).
\bibitem{Sjoberg} O.\ Sj\"oberg, Nucl.\ Phys.\ A {\bf 265}, 511 (1976).
\bibitem{Pilati} S.\ Pilati and S.\ Giorgini, Phys.\ Rev.\ Lett.\
{\bf 100}, 030401 (2008).
\bibitem{KB3G} A.\ Poves, J.\ S\'anchez-Solano, E.\ Caurier and F.\
Nowacki, Nucl.\ Phys.\ A {\bf 694}, 157 (2001).
\bibitem{GXPF1} M.\ Honma, T.\ Otsuka, B.\ A.\ Brown and T.\ Mizusaki,
Phys.\ Rev.\ C {\bf 69}, 034335 (2004).
\bibitem{AME2003} G.\ Audi, A.\ H.\ Wapstra and C.\ Thibault,
Nucl.\ Phys.\ A {\bf 729}, 337 (2003).
\bibitem{53-58Ca} M.\ Langevin {\it et al.}, Phys.\ Lett.\ B {\bf 130},
251 (1983); M.\ Bernas {\it et al.}, Phys.\ Lett.\ B {\bf 415}, 111
(1997); O.\ B.\ Tarasov {\it et al.}, Phys.\ Rev.\ C {\bf 80}, 
034609 (2009).
\bibitem{coreNCSM} A.\ F.\ Lisetskiy, B.\ R.\ Barrett, M.\ K.\ G.\ Kruse,
P.\ Navr\'{a}til, I.\ Stetcu and J.\ P.\ Vary, Phys.\ Rev.\ C {\bf 78}, 
044302 (2008).
\bibitem{CC} G.\ Hagen, T.\ Papenbrock, D.\ J.\ Dean and M.\ Hjorth-Jensen, 
Phys.\ Rev.\ C {\bf 82}, 034330 (2010).
\bibitem{lattice} E.\ Epelbaum, H.\ Krebs, D.\ Lee and U.-G.\ Mei{\ss}ner,
 Eur.\ Phys.\ J.\ A {\bf 45}, 335 (2010).
\bibitem{CC3N} G.\ Hagen, T.\ Papenbrock, D.\ J.\ Dean, A.\ Schwenk, A.\ 
Nogga, M.\ W{\l}och and P.\ Piecuch, Phys.\ Rev.\ C {\bf 76}, 034302 (2007).
\bibitem{IMSRG} K.\ Tsukiyama, S.\ K.\ Bogner and A.\ Schwenk,
arXiv:1006.3639.
\bibitem{chiralnm} K.\ Hebeler and A.\ Schwenk, Phys.\ Rev.\ C {\bf 82},
014314 (2010).
\end{thebibliography}
\end{document}